# Thermal and electrical signatures of a hydrodynamic electron fluid in WP$_2$

J. Gooth[1*†], F. Menges[1$], N. Kumar[2], V. Süβ[2], C. Shekhar[2], Y. Sun[2], U. Drechsler[1], R. Zierold[3], C. Felser[2], B. Gotsmann[1*]

[1]*IBM Research - Zurich, Säumerstrasse 4, 8803 Rüschlikon, Switzerland.*

[2]*Max Planck Institute for Chemical Physics of Solids, Nöthnitzer Straße 40, 01187 Dresden, Germany.*

[3]*Institute of Nanostructure and Solid-State Physics, Universität Hamburg, Jungiusstraße 11, 20355 Hamburg, Germany*

*johannes.gooth@cpfs.mpg.de, bgo@zurich.ibm.com*

[$] *now at University of Colorado Boulder, Department of Physics, Boulder, CO, USA*

[†] *now at Max Planck Institute for Chemical Physics of Solids, Nöthnitzer Straße 40, 01187 Dresden, Germany*

**Materials with strongly-interacting electrons exhibit interesting phenomena such as metal-insulator transitions and high-temperature superconductivity.[1] In stark contrast to ordinary metals, electron transport in these materials is thought to resemble the flow of viscous fluids.[2] Despite their differences, it is predicted that transport in both conventional and correlated materials is fundamentally limited by the uncertainty principle applied to energy dissipation.[3–5] Here we report on the observation of experimental signatures of hydrodynamic electron flow in the Weyl semimetal tungsten diphosphide ($WP_2$). Using thermal and magneto-electric transport experiments, we find indications of the transition from a conventional metallic state at higher temperatures to a hydrodynamic electron fluid below 20 K. The hydrodynamic regime is characterized by a viscosity-induced dependence of the electrical resistivity on the square of the channel width and by a strong violation of the Wiedemann–Franz law. From magneto-hydrodynamic experiments and complementary Hall measurements, the relaxation times for momentum relaxing and conserving processes are extracted. Following the uncertainty principle, both electrical and thermal transport are limited by the Planckian bound of dissipation, independent of the underlying transport regime.**

In an overwhelmingly large group of conducting materials, charge transport can be described by the rather simple model of a free-electron gas. Its basis is that the carriers move unimpededly until they scatter with phonons or defects. Such collisions usually relax both the momentum and the energy currents, and consequently impose a resistance to charge and heat flow alike. In most conventional conductors, electrical and thermal transport are therefore related *via* the Wiedemann–Franz law, which states that the product of the electrical resistivity $\rho$ and the thermal conductivity $\kappa$, divided by the temperature $T$ is a constant $L = \rho\kappa/T$, yielding the

Sommerfeld value $L_0$ = 2.44×10⁻⁸ WΩK⁻². $\rho$ and $\kappa$ are intrinsic material properties and independent of the size and geometry of the conducting bulk. However, the conventional free-electron model fails to describe transport in strongly-interacting electron systems.[2] The difficulty is to find a theoretical framework that captures the frequent inter-particle collisions that define the interaction within the many-body system. Recently, it has been rediscovered that the theory of hydrodynamics, which is normally applied to explain the behavior of classical liquids like water, could be used to describe the collective motion of electrons in such a system [6,3,7–9].

In contrast to the free-electron gas, the energy dissipation in a hydrodynamic electron fluid is dominated by momentum-conserving electron-electron scattering or small-angle electron-phonon scattering. A signature of the hydrodynamic nature of transport emerges when the flow of the electrons is restricted to channels [10,11]. The electrical resistance of a hydrodynamic electron liquid is then proportional to its shear viscosity, and therefore paradoxically increases with increasing mean free path of the electrons [6,12]. Viscosity-induced shear forces at the channel walls cause a nonuniform velocity profile, so that the electrical resistivity becomes a function of the channel width. Moreover, the electrical resistivity will become small with increasing width, because momentum-relaxing processes within the bulk are strongly suppressed. The thermal conductivity is instead dominated by faster momentum-conserving collisions, Consequently, a strong violation of the Wiedemann–Franz law is predicted [3,7,8,13].

Despite the significant difference in the microscopic mechanisms behind momentum- and energy-current-relaxing collisions, it has recently been proposed that dissipation in both processes is tied to a universal upper bound of entropy production with a time scale of $\tau_\hbar = \hbar/(k_B T)$ [1,3–5], where $\tau_\hbar$ is determined only by the Boltzmann constant $k_B$, the reduced Planck constant $\hbar$ and the temperature $T$. This concept of "Planckian dissipation" has been developed in the frameworks of the uncertainty principle [3] and of a string theory, known as anti-de-Sitter space/conformal field theory correspondence (AdS/CFT) [2,14]. Holographic models successfully predicted the universal bound on the momentum-relaxation time $\tau_{\text{mr}}$ in a strongly interacting neutral plasma [15]. The momentum-relaxation bound $\tau_{\text{mr}} \geq \tau_\hbar$ can also be expressed as the ratio of the shear viscosity to the entropy density, and is not only supported by experiments on the quark-gluon plasma and on the ultra-cold Fermi gas [16,17], but is also well-respected in classical fluids such as water.

Likewise, maximally dissipative processes matching the timescale $\tau_\hbar$ have recently been proposed to underpin the *T*-proportional resistivity of metals [3,4], and are believed to be at the root of high-temperature superconductivity [5,18]. In the context of charge transport described by quasi-particles, $\tau_{\text{mr}}$ represents the characteristic scattering time to randomize the excess forward momentum of a quasi-particle. Momentum relaxation in most conductors is determined by the way in which the electrons couple *via* Umklapp processes to the lattice or to the disorder of the host solid. Thus, $\tau_{\text{mr}}$ of the electron system is determined by extrinsic coupling parameters and is not generally universal [3]. In the hydrodynamic regime, the momentum and energy-current relaxations are independent processes, which in principle enables the isolation of the momentum-conserving time relaxation time $\tau_{\text{mc}}$. The characteristic time $\tau_{\text{er}}$ needed to dissipate the excess energy of a quasi-particle includes both momentum-relaxing and conserving scattering processes. As such, to a first approximation the momentum-conserving scattering time is $1/\tau_{\text{er}} = 1/\tau_{\text{mc}} + 1/\tau_{\text{mr}}$ [8].

Hydrodynamic effects have been postulated to play a role in the *T*-linear resistivity of high-temperature superconductors above their critical temperature, but purely hydrodynamic transport is not directly applicable to most of those systems. Extracting $\tau_{\text{mr}}$ and $\tau_{\text{er}}$ separately from experimental data has remained challenging because of the strong momentum-relaxation contribution. In fact, as momentum-relaxation processes are always present in a real material system, momentum can only be quasi-conserved. This, however, does not mean that hydrodynamic signatures are not observable in transport experiments. Hydrodynamic effects become dominant, when the momentum-conserving scattering length $l_{mc} = v_F \tau_{mc}$ provides the smallest spatial scale in the system, $l_{mc} \ll w \ll l_{mr}$, where $w$ is the sample width, $l_{mr} = v_F \tau_{mr}$ the momentum-relaxing scattering length and $v_F$ the Fermi velocity [10,19].

Inspired by pioneering experiments on semiconductor wires [20], signatures for hydrodynamic electron flow were recently reported in ultraclean $PdCoO_2$ [19] and graphene [21–23]. However, as these materials exhibit only relatively weak scattering, experimental evidence of a universal thermal dissipation bound has been elusive. It is therefore desirable to go beyond previous experiments and investigate the dissipative timescales of a strongly-correlated material in the limit of hydrodynamics.

For our study, we have chosen the semimetal $WP_2$ [24] because momentum-relaxing Umklapp scattering is anomalously strongly suppressed in this material. $WP_2$ exhibits a space-group symmetry with two pairs of two-fold degenerated Weyl points close to the intrinsic Fermi level. $WP_2$ contains a mirror plane perpendicular to the *a*-axis, a *c*-glide perpendicular to the *b*-axis and a two-fold screw axis along the *c*-axis. Owing to the high crystalline anisotropy, typical $WP_2$ crystals are needle-shaped with an orientation along the *a*-axis. Moreover, the magneto-transport in bulk single crystals has been shown to be highly anisotropic by 2.5 orders of magnitudes between the *a-c* and the *a-b* plane (Extended Data Fig. 1), similarly to bulk $PdCoO_2$ [25]. We note that although $WP_2$ is referred to as a semimetal, it exhibits a finite density of free charge carriers at the Fermi energy (Extended Data Fig. 2). The best single-crystalline bulk samples of $WP_2$ exhibit a *T*-linear electrical resistivity above 150 K that is dominated by electron-phonon scattering and a temperature-independent resistivity below 20 K as previously observed in PdCoO$_2$ [19]. Its residual resistivity is only 3 nΩcm, i.e., a remarkable three times lower value than that of PdCoO$_2$ in the hydrodynamic regime. At temperatures between 20 and 150 K, $\rho$ increases exponentially with increasing *T*, which has been attributed to the phonon-drag effect. Phonon drag is considered beneficial for reaching the hydrodynamic regime, because it provides another source of momentum-conserving scattering [19]. At 4 K, the bulk samples exhibit a mean free path of $l_{mr} \approx 100$ μm in the *a-c* direction (see Methods for details). The $l_{mr}$ of WP$_2$ exceeds the momentum-relaxing scattering length of hydrodynamic PdCoO$_2$ and that of graphene [21,22] by one and two orders of magnitude, respectively. These properties make WP$_2$ an ideal material for investigating hydrodynamic effects and the associated dissipative bounds in its strongly-correlated electron system.

For our experiments, we produced a series of WP$_2$ micro-ribbons by milling chemical vapor transport grown single crystals. The high anisotropy of the crystals results in rectangular micro-samples that retain the needle-shaped orientation of the grown crystal. Using a micro-manipulator, the milled micro-ribbons were transferred to a pre-defined metallic line structure (Fig. 1 (a)). The solely mechanical fabrication method prevented chemical contamination and damage of the ultra-pure source material. Electron-beam deposition of platinum was used to provide electrical contacts by connecting the ribbon ends to the underlying metal lines. Electron microscopy was used to determine the distance between the contact lines, i.e., the length *l* of the micro-ribbon along the *a*-axis of the crystal, its thickness *t* along the *b*-axis, and its width *w* along the *c*-axis. The transport direction in our samples matches the crystal's *a*-axis. The thickness is approximately $t = w/2$. However, the high anisotropy of the magneto-transport yields a mean free path in the *a-b* direction that is about 250 times lower than that in the *a-c* direction [25]. Therefore, *w* is the characteristic length scale of the samples [24], justifying the use of two-dimensional models for the in-plane transport properties in *a-c*. We investigated four micro-ribbons with widths of 0.4 µm, 2.5 µm, 5.6 µm and 9.0 µm, all satisfying $w \ll l_{mr}$ at low temperatures. Note that to observe the surface effects of all walls in equal strength in the transport experiments, one would have to design the sample such that its width is 250 times larger than its height.

In a first set of transport experiments, we studied the *T*- and *w*-dependence of the electrical resistivity $\rho = V/I \cdot wt/l$ of the micro-ribbons (Fig. 1 (b)). For this purpose, we measured the voltage response *V* to an AC-current bias *I*, using the standard low-frequency lock-in technique (see Methods for details). The elongated geometry of the micro-ribbons with contact lines wrapping around the whole cross section of the samples was chosen to ensure homogenous current distributions. Because of the low resistivity of the bulk sample, special care must be taken in extracting the intrinsic $\rho$ of the WP$_2$ ribbons. We therefore compare four-terminal with quasi-four-terminal resistivity measurements. The quasi-four-terminal measurements exclude the resistance of the contact lines, but in principle include the interface resistances at the metal/semimetal junction. The electrical contact resistances are found, however, to be negligibly small, and not measurable in our experiment (see Fig 1 (d) and the Methods for a detailed analysis). Conventionally, $\rho = \rho_0 = m^*/(e^2 n \tau_{mr})$ is an intrinsic bulk property and does not depend on *w*. According to the Drude model, $\rho$ only depends on the effective mass $m^*$ of the charge carriers, the elementary charge *e* and the carrier concentration *n*. However, when boundary scattering becomes significant, $\rho$ can turn into a function of the sample size [23]. In the well-established ballistic regime ($w \ll l_{er}, l_{mr}$), for example, the electrical resistivity is given by $\rho \sim w^-$

[1]. Further, in a hydrodynamic fluid ($l_{er} \ll w \ll l_{mr}$), the flow resistance is determined solely by the interaction with the sample boundaries, reducing the average flow velocity of the electron fluid (Fig. 1 (c)). Recent theory predicts that the electrical resistivity in the Navier–Stokes flow limit is modified as $\rho = m^*/(e^2 n) \cdot 12\eta w^{-2}$, where electron-electron and small-angle electron-phonon scattering are parameterized in the shear viscosity $\eta$ (6–8). Therefore, different transport regimes are characterized by the exponent $\beta$ of the width dependence, $\rho \sim w^\beta$.

As shown in Fig. 1 (b), all ribbons investigated consistently exhibit a constant $\rho = \rho_0$ above 150 K. In accordance with the bulk measurements [24], $\rho$ increases linearly with increasing $T$, as expected for dominant electron-phonon scattering. At lower temperatures, however, $\rho$ becomes a non-monotonic function of $T$ and increases with decreasing $w$. The change of slope in $\rho(T)$ is more pronounced in narrower ribbons, corroborating the importance of the sample's spatial boundaries in this temperature regime. As in real materials the momentum is only quasi-conserved, $\rho$ always contains a width-independent Drude offset $\rho_0$ in addition to the width-dependent power-law component $\rho_1 w^\beta$. To extract the power $\beta$ from the experimental data (Fig. 1 (d) and Extended Data Fig. 5), we have subtracted $\rho_0$ from $\rho$ at all temperatures, fitting the experimental data with $\rho = \rho_0 + \rho_1 w^\beta$. The exponents obtained were then cross-checked by a logarithmic analysis of $\rho - \rho_0 = \rho_1 w^\beta$ (Extended Data Fig. 8). As shown in Fig. 1 (d), we found that the low-temperature regime ($T < 20$ K) is well described by an inverse quadratic relation $\rho - \rho_0 \sim w^{-2}$, in agreement with the Navier–Stokes description of hydrodynamic flow [10–12]. The residual resistivity $\rho_0$ of about 4 n$\Omega$cm obtained at 4 K matches the bulk resistivity excellently. This result enables a quantitative extraction of the kinematic shear viscosity of the electron liquid as $\eta = 3.8 \times 10^{-2}$ m$^2$s$^{-1}$ at 4 K (see Methods for details). Multiplication by the mass density $M =$

$nm^*$ yields a dynamic viscosity of about $\eta_D = 1\times10^{-4}$ kgm$^{-1}$s$^{-1}$ at 4 K, which is on the order of that of liquid nitrogen at 75 K. The ratio between the dynamic viscosity and the number density is therefore 430 $\hbar$. The observed $w^{-2}$ dependence provides strong evidence of hydrodynamic electron flow in WP$_2$, whereas the regime between 20 K and 150 K can be explained by a smooth transition to a hybrid state in which viscosity-stimulated boundary scattering mixes with momentum-relaxing electron-phonon collisions [19].

Next, we investigated the Lorenz number $L = \kappa\rho/T$ in WP$_2$, which is widely considered to be an important observable for characterizing thermal and charge transport characteristics [8,13,22,26]. Therefore, we determined the thermal conductivity $\kappa$ of the 2.5-µm-wide WP$_2$ sample. The measurements were performed with open boundary conditions, prohibiting electric current flow. Zero electric current forces zero momentum flux, which decouples the heat flow from the momentum drag in the hydrodynamic regime [13]. The ribbon was mounted on a microsystem platform [26,27] with two integrated heater/sensors as shown in Fig. 2 (a). The sensor device is thermally insulated through 1.2-mm-long silicon nitride bars operated in vacuum to $1.6\times10^{-5}$ K/W at room temperature. The fabrication and characterization of the platform are described in detail in the Supplementary Materials. Two gold resistors, serving as both thermometers and heaters, were calibrated at each temperature and used to measure both the temperature bias along and the heat flux through the WP$_2$ sample. Although used routinely for the thermal characterization of micro- and nanoscale samples, this method often suffers from thermal contact resistance effects, in particular when applied to the characterization of nanoscale structures. To minimize such effects, we chose large dimensions for both the sample as well as for the electrical contacts. Nevertheless, we calculate relatively large expected systematic errors as shown as error bars in the plot. The experimentally extracted $\kappa$ as a function of $T$ is given in Fig. 2 (b).

As shown in Fig. 2 (c), the Wiedemann–Franz relationship holds above 150 K, as the Lorenz number assumes the Sommerfeld value $L \approx L_0$. With decreasing temperature, however, $L$ is strongly reduced by more than one order of magnitude to a value below 0.05 $L_0$ at 20 K. We note, that our thermal conductance measurements are in excellent agreement with independent measurements on macroscopic WP$_2$ crystals leading to $L \approx 0.1\ L_0$ at $T = 15$ K, despite the full phonon contribution being included.[28]

To interpret, we recall that in conventional conductors, the Wiedemann–Franz law holds, i.e., $L$ equals the Sommerfeld value $L_0 = \pi^2\ k_B^2/(3e^2)$. Violations of the Wiedemann–Franz law typically are an indication of invalidity of the quasi-particle picture, strong difference between $\tau_{er}$ and $\tau_{mr}$ in quasi-particle scattering, or ambipolar physics. Phonon contributions to $\kappa$ enhance $L$. Conversely, small ($O(1)$) deviations below $L_0$ are usually observed in metals near 0.1 of the Debye temperature. Ultra-pure metals at low temperatures, for example, can reach of $L/L_0 \sim 0.1$ in extreme cases. This effect is due to electron - phonon scattering and a transition from inelastic large angle to small angle scattering processes.[12,29] In hydrodynamic systems, $L$ can become arbitrarily small because of the difference between the two relaxation times $\tau_{mr}$ and $\tau_{er}$ governing electrical and thermal transport, respectively [3,8,13,30]. Hence, we expect the ratio between $\tau_{mr}$ and $\tau_{er}$ in WP$_2$ to be at least one order of magnitude [8]. However, we note that this can only be a lower bound of the difference between the scattering times in the electron system because the residual phonon contributions are not subtracted from $\kappa$. The observed maximum in thermal conductance at around 10 K is consistent with this notion. Nevertheless, the consideration of a separate heat-conduction channel *via* the crystal lattice through phonons must be made with care in correlated materials. A recent claim argues that the hydrodynamic fluid may comprise both phonons and electrons and it is sometimes referred to as an "electron phonon soup" [31]. In any case, the measured thermal conductance shown here includes both phonon and electron contributions.

The Lorenz value we obtained belongs to the lowest such values ever reported [26]. This is an indication for strong inelastic scattering, and, in combination with the charge transport data shown above, an independent evidence of a hydrodynamic electron fluid below 20 K.

Exploiting the well-justified conjecture of the hydrodynamic nature of transport in WP$_2$ at low temperatures, we can now manipulate and tune the viscosity of the electron fluid to obtain further information on the magnitude of $\tau_{mc}$. For this, the resistivity of the micro-ribbons is measured as a function of the magnetic field $\boldsymbol{B}$ at fixed $T$. $|\boldsymbol{B}| = B$ is set along the $b$-axis of the crystal and thus perpendicular to the direction of current flow. In an electron liquid, the viscosity is defined by the internal friction between layers of different velocities [10], mediated by the exchange of electrons (Fig. 3 (a)). The strength of the friction is given by the penetration depth $\lambda_p$ of the electrons, which is on the order of the mean free path between collisions, $\lambda_p \approx l_{mc}$, at zero magnetic field. When $\boldsymbol{B}$ is turned on, however, this penetration depth is limited by the cyclotron radius $r_c = mv_F/eB$. Thus, in a strong magnetic field, the viscosity should tend to zero, providing a mechanism for a large negative magneto-resistivity. Solving the corresponding magnetohydrodynamic equations results in the magnetic-field-dependent viscosity along the $a$-axis of the crystal $\eta(B) = \eta_0/(1 + (2\tau_{mc}\omega_c)^2)$, where $\omega_c = v_F/r_c$ is the cyclotron frequency [10,11,29].

As shown in Fig. 3 (b), we observe a large negative magneto-resistivity at low temperatures. When $\rho(B)-\rho_0$ is normalized by $w^{-2}$, all experimental data below 20 K collapse onto a single curve, matching the magnetic-field-dependent viscosity model excellently. The average viscosity of the single traces at zero field $\eta_0$ equals the viscosity $\eta$ obtained from the $w^{-2}$ fits above. This agreement is an important cross-check, confirming our results and the hydrodynamic interpretation. Furthermore, in accordance with the theory, $\eta(B)$ tends to zero as the magnetic field is enhanced.

The $\tau_{mc}$ extracted at individual $T$ values are plotted in Fig. 3 (c). Throughout the hydrodynamic regime, we find that the momentum-conserving scattering time in the electron liquid is tied to the Planckian limit [1,3,5] as $\tau_{mc} \sim \tau_\hbar = \hbar/(k_B T)$. As an important cross-check, we calculated the kinematic shear viscosity [10,11] from $\tau_{mc}$ as $\eta = v_F^2 \tau_{mc}/4$, and obtained consistent values with those extracted above (Extended Data Fig. 10), confirming the underlying dissipation bound.

For comparison, we also determined the timescale of the charge current relaxation related to the electrons' momentum. $\tau_{mr}$ is extracted from combined resistivity and Hall measurements on the bulk sample [24] (Extended Data Fig. 2 (a)). With the carrier concentration $n$ (Extended Data Fig. 2 (b)) obtained from Hall measurements and Shubnikov–de Haas oscillations [24], we calculated the mobility $\mu = (\rho e n)^{-1}$ (Extended Data Fig. 2 (c)). Using the average effective mass of the charge carriers $m^* = 1.21\, m_0$ ($m_0$ is the free electron mass) and $\mu = e\tau_{mr}/m^*$, we eventually obtained $\tau_{mr}$ as a function of the temperature (Fig. 3 (c)). At low temperatures, $\tau_{mr}$ is three orders of magnitudes larger than $\tau_{mc}$ and thus also than the Planckian bound. Consequently, $l_{mc} < w \ll l_{mr}$ (Extended Data Fig. 11), validating the hydrodynamic treatment of transport in WP$_2$ in this temperature regime.

A natural question that arises from this observation is whether Planckian dissipation is exclusive to the hydrodynamic regime in WP$_2$. In the $T$-linear regime above 150 K (Extended Data Fig. 3), where strong electron-phonon interactions yield $L = L_0$ (because $\tau_{mc} = \tau_{mr}$), we found that, despite their fundamental difference, electron-phonon and electron-defect processes are also tied to $\tau_\hbar$. This observation indicates that the dissipation in WP$_2$ is generally tied to a characteristic time scale $\tau_\hbar$, regardless of the transport regime and the details of the underlying scattering mechanisms.

Note that our analysis relies on the picture of weakly-interacting quasiparticles. The consistency between the different analysis steps appears to confirm this notion, which largely builds on Fermi liquid theory. This is on contrast to previous inferring of the Planckian bound to describe dissipation, which exhibit either strongly interacting electron systems or strong electron-phonon coupling above the Debye temperature. Our findings therefore underline the wide applicability of the Planckian bound.

In conclusion, our experiments strongly support the existence of a hydrodynamic electron fluid in WP$_2$. The accompanying independence of $\tau_{mc}$ and $\tau_{mr}$ allows the intrinsic thermal current relaxation process to be isolated, which is particularly elusive in other contexts. Remarkably, it turns out that the electron system in WP$_2$ generates entropy in a very simple and universal way in which the only relevant scale is the temperature.

# FIGURES

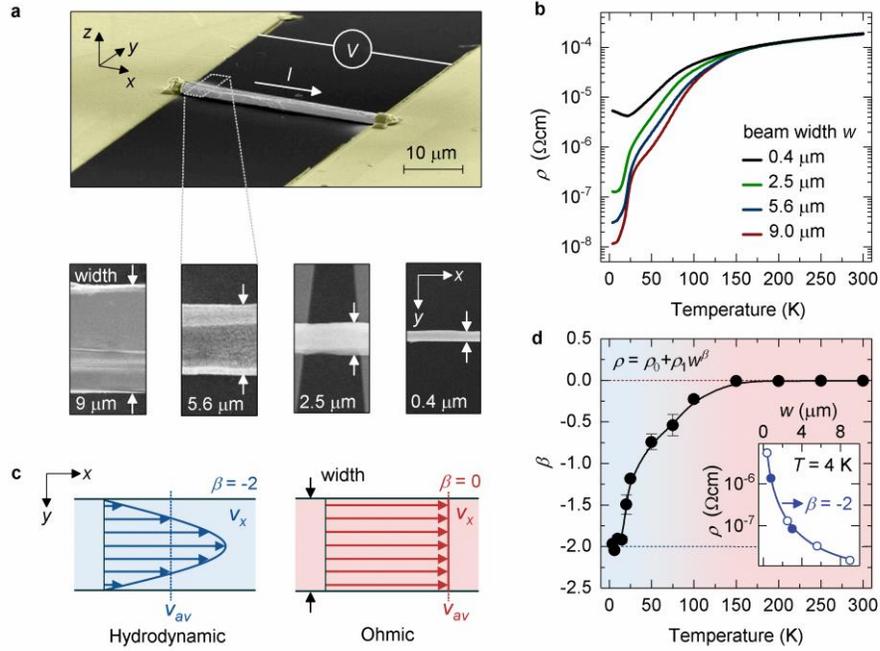

**Figure 1 | Effect of the channel width on the electrical resistivity. a**, False-colored scanning electron-beam microscopy (SEM) image of the device to measure the electrical resistance $R = V/I$ (upper panel) of the $WP_2$ micro-ribbons. Ribbons of four different widths $w$ were investigated (lower panels). The error of the measured width is below 5%, including the uncertainty of the SEM and sample roughness. **b**, Temperature-dependent electrical resistivity $\rho$ of the four ribbons. **c**, Sketch of the velocity $v_x$ flow profile for a Stokes flow (left panel) and conventional charge flow (right panel). $v_{av}$ is the average velocity of the charge-carrier system. **d**, Exponent $\beta$ of the functional dependence $\rho(w)$ as a function of temperature, extracted from power law fits $\rho = \rho_0 + \rho_1 w^\beta$ of the data plotted in (b). The inset shows an exemplarily power-law fits at 4 K, where the open and filled symbols represent quasi-four and four terminal measurements, respectively. The line is a power law fit, leading to $\beta = -2$. The colored

background marks the hydrodynamic (light blue) and the normal metallic (Ohmic) temperature regime (light red).

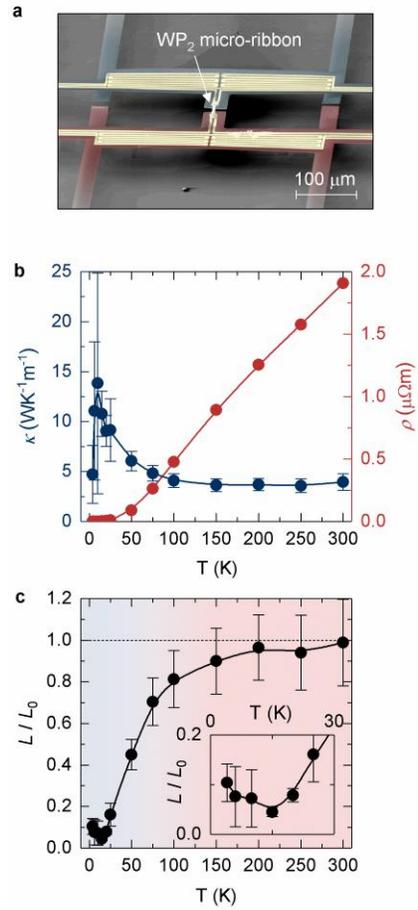

**Figure 2 | Violation of the Wiedemann–Franz law. a**, False-colored SEM image of a microdevice for measuring thermal transport that consist of two suspended platforms bridged by a 2.5-µm-wide $WP_2$ ribbon. **b**, Thermal conductivity $\kappa$ (left axis) and electrical resistivity $\rho$ (right axis) of the micro-ribbon as a function of temperature. **c**, Lorenz number $L = \kappa T \rho$, calculated from the data in (b). The inset shows a zoom of the low-temperature region.

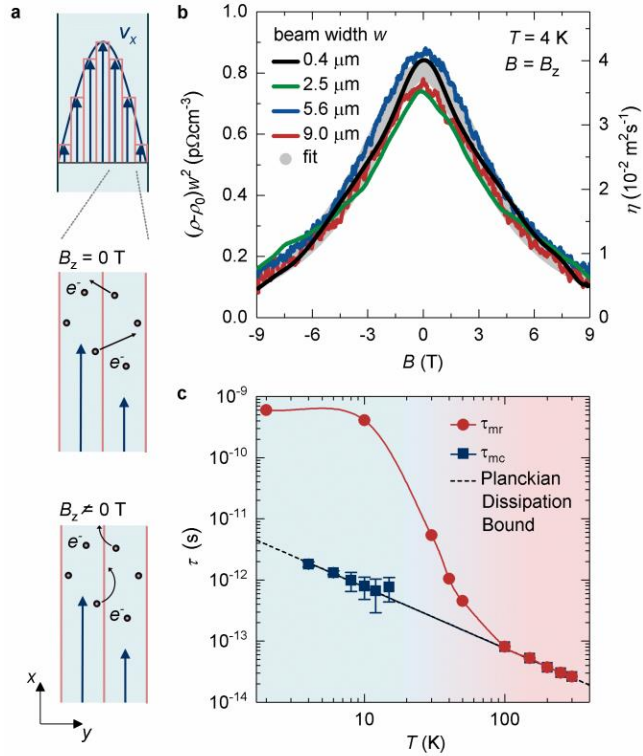

**Figure 3 | Magnetohydrodynamics and the Planckian bound of dissipation. a**, The origin of the decrease of the electron viscosity $\eta$ in a magnetic field $B$ (schematic) perpendicular to the current flow and the sample width. The viscous friction between two adjacent layers of the electron fluid moving with different velocities is determined by the depth of the interlayer penetration of the charge carriers $e^-$. In a magnetic field, this depth is limited by the cyclotron radius. **b**, $(\rho-\rho_0)/w^2$ as a function of $B$ for all four WP$_2$ ribbons investigated at 4 K (lines). The experimental data has been fitted by the magnetohydrodynamic model in the Navier–Stokes flow limit (grey dots). **c**, Experimentally extracted momentum-relaxing and momentum-conserving relaxation times $\tau_{mr}$ and $\tau_{mc}$, respectively (symbols, with guide to the eye). The dashed line marks the Planckian bound on the dissipation time $\tau_\hbar = \hbar/(k_B T)$.

# METHODS

## WP$_2$ single-crystal growth of the bulk sample

Crystals of WP$_2$ were prepared by chemical vapour transport method. Starting materials were red phosphorous (Alfa-Aesar, 99.999 %) and tungsten trioxide (Alfa-Aesar, 99.998 %) with iodine as a transport agent. The materials were taken in an evacuated fused silica ampoule. The transport reaction was carried out in a two-zone-furnace with a temperature gradient of 1000 °C to 900 °C for serval weeks. After reaction, the ampoule was removed from the furnace and quenched in water. The metallic-needle crystals were characterized by X-ray diffraction.

## Electrical Transport Measurements on the Bulk Sample

The electrical resistance of the bulk sample is determined within a four-probe configuration under isothermal conditions with an AC bias current of r.m.s. 3 mA at 93.0 Hz, using standard lock-in technique. As for the microbeams, the current is applied along the *a*-axis of the crystal. Magnetoresistance measurements are performed with a magnetic field applied perpendicular to the direction of current flow, within the *b-c* plane. The Hall resistance is measured under the same conditions with a magnetic field applied along the *b*-axis. More details on the bulk characterization can be found in Kumar *et al.*[24]. All measurements are carried out in a temperature variable cryostat, equipped with a superconducting ±9 T magnet (PPMS Dynacool). Magnetic field-dependent electrical resistance measurements on the WP$_2$ bulk sample, revealed an enormous, highly anisotropic magneto-resistance (MR) (Extended Data Fig. 1)[24]. The anisotropy has been attributed to orbital motions of charge carriers on the anisotropic Fermi surfaces, driven by the Lorentz force. Such effect has previously been reported[25] in PdCoO$_2$, which becomes hydrodynamic at low temperatures[19]. The anisotropy of the Fermi surface manifests itself in the Fermi velocity $v_F$, which is directly related to the mean free path $l_{mr}$ in the different crystal planes perpendicular to the magnetic field. The MR is, therefore, maximum when the field is applied along the *b*-axis of the crystal (Extended Data Fig. 1) and is decreased by 2.5 orders of magnitudes when the field is rotated to the *c*-axis. Consequently, the $l_{mr}$ in the *a-c* plane is 250-times larger than in the *a-b* plane.

From the Hall (Extended Data Fig. 2 (a)) measurements [24], we obtain the temperature (*T*)-dependent average carrier concentration $n = (d\rho_{xy}/dB\ e)^{-1}$ above 30 K from a single band model. Below 30 K, the Hall resistivity becomes non-linear, indicating the contributions of multiple bands to the transport. At these low temperatures, four frequencies *f* were determined from Shubnikov-de Haas oscillations when the magnetic field is applied along the *b*-axis, of which two are hole-pockets $\alpha$ and $\beta$ ($f_\alpha = 1460$ T and $f_\beta = 1950$ T) and two are electron-pockets $\gamma$ and $\delta$ ($f_\gamma = 2650$ T and $f_\delta = 3790$ T) [24]. These pockets were identified using the modified Becke-Johnson method (Extended Data Fig. 2). We calculated the k-space areas of the extremal cross-sections of the Fermi surfaces perpendicular to a magnetic field that is applied along the *b*-axis of the crystal and found four frequencies at around 1300 T, 1900 T, 2800 T and 3900 T, which reflect the experimentally obtained frequencies well. We have recently determined the effective masses of the hole-pockets from the Shubnikov de-Haas oscillations as $m_\alpha = 1.67\ m_0$ and $m_\beta = 1.89\ m_0$. To account for all bands at the Fermi level, we additionally calculated the effective masses of the electron-pockets as $m_\gamma = 0.87\ m_0$ and $m_\delta = 0.99\ m_0$, where $m_0$ denotes the free electron mass (Extended Data Fig. 3).

From the Onsager relation, we determine the size of the Fermi surface cross sections $A_{Fi} = 2\pi^2 f_i /\phi_0$ with $\phi_0$ as the magnetic flux quantum and the sub-index *i* reflecting $\alpha$, $\beta$, $\gamma$ and $\delta$. Applying the standard circular approximation, we obtain momentum-vectors $k_{Fi} = (A_{Fi}/\pi)^{1/2}$ of $k_{F\alpha} = 2.11$

$\times 10^9$ m$^{-1}$, $k_{F\beta} = 2.40 \times 10^9$ m$^{-1}$, $k_{F\gamma} = 2.84 \times 10^9$ m$^{-1}$ and $k_{F\delta} = 3.39 \times 10^9$ m$^{-1}$. Fermi liquid theory in the limit of zero temperature results in the total carrier concentration of $n = \sum n_i = 2.5 \times 10^{21}$ cm$^{-3}$, where $n_i = k_{Fi}^3/(3\pi^2)$. Given the carrier concentrations for the whole temperature range investigated (Extended Data Fig. 2 (b)), we now calculate the average mobility $\mu = (\rho e n)^{-1}$ of the WP$_2$ bulk sample (Extended Data Fig. 2 (c)). The corresponding Fermi velocities $v_{Fi} = \hbar k_{Fi}/m_i$ are $v_{F\alpha} = 1.89 \times 10^5$ m/s, $v_{F\beta} = 1.67 \times 10^5$ m/s, $v_{F\gamma} = 3.88 \times 10^5$ m/s and $v_{F\delta} = 3.19 \times 10^5$ m/s. To account for all band contributions in the electrical transport, we calculated the harmonic mean giving the average effective mass $m^* = 4(1/m_\alpha + 1/m_\beta + 1/m_\gamma + 1/m_\delta)^{-1} = 1.21\ m_0$ and the mean Fermi velocity $v_F = (\sum v_{Fi})/4 = 2.59 \times 10^5$ m/s. The $l_{mr}$ in the $a$-$c$ plane is then determined as a function of temperature from the $T$-dependent mobility from as $\mu = e v_F l_{mr} / m^*$.

**Electrical Transport Measurements on the Microbeams**

All transport measurements are performed in a temperature-variable cryostat (Dynacool, Quantum Design) in vacuum. The cryostat is equipped with a ±9 T superconducting magnet, swept with a rate of 5 mT/s. After fabrication, the microbeam devices are mounted on a sample holder and wire bonded. Electrical resistance measurements on the microbeams are carried out in a quasi-four-probe configuration under isothermal conditions with AC bias currents of r.m.s. 100 µA max at 6.1 Hz, using standard lock-in technique. The quasi-four terminal resistivity measurements exclude the resistance of the contact lines, but can in principle include interface resistances at the metal/semimetal junctions. However, we find no evidence for any resistance contribution from this interface (see following paragraph).

**Interface resistance at the metal/semimetal junction**

The interface resistance is an important issue, dealing with low sample electrical resistances in the range of mΩ. We therefore have carried out three independent cross-checks to evaluate interface resistance in our devices. All three methods show independently that the interface resistance at the metal/semimetal junction is negligible small in our experiment, despite the low sample resistance:

1. A common method to evaluate the contact resistance is to compare four 4- and 2-point (or quasi-4-point) measurements. Quasi-4-terminal (red) and actual 4-terminal (blue) resistivity of the 9 μm-wide $WP_2$ ribbon as a function of temperature are shown in Extended Data Fig. 4. The deviation between the two curves is below 1 % at 300 K and rises to 10 % at 4 K. We note, however, that this enhancement at low temperature is within the measurement error, due to the low resistivity in this temperature range. The 9 μm-wide has the lowest resistance and should therefore be most sensitive to the interface resistance.

2. Further, we have used a Focused Ion Beam to cut out four-terminal devices. Two devices of 0.8 μm and 3.5 μm width have been fabricated (Extended Data Fig. 5). The resistivity data points obtained from these devices fit in perfectly to the width-dependent series.

3. Next, we have checked the residual resistivity $\rho_0$ of the width-dependent resistivity, fitting the data by $\rho = \rho_0 + \rho_1 w^\beta$. As shown exemplarily Extended Data Fig. 5 for the 4 K data, we find $\rho_0$ matching the resistivity values of the bulk samples measured in Kumar et al.[24]. Explicitly, $\rho_0(4\ K) = 4.1 \pm 0.4$ nΩcm.

4. An important cross-check that the evaluated $\rho_0$ itself does not introduce the width dependence is to determine $\rho_0$ from the magnetic field-dependent resistivity data of each width independently. We therefore fit each magnetoresistance curve individually (Extended Data Fig. 6) by the $\rho = \rho_0 + \rho_{1,a} w^{\beta}/(1+(\rho_{1,b}B)^2)$, where $B$ is the magnetic field and $\beta$ is extracted from the width-dependent analysis above. As shown in Extended Data Fig. 7, all individual curves result in a $\rho_0(4\text{ K})$ of around 4, in full agreement with the previous analysis.

Therefore, we conclude that within our measurement precision, the metal/semimetal interface perfectly transmits charge carriers and does not represent a noticeable resistance. This could indicate a different mechanism for resistances at hydrodynamic/normal metal electron interfaces.

**Exponent of the functional dependence of $\rho$ on $w$**

Because momentum-relaxation processes are always present in the bulk of the investigated material system, momentum can only be quasi-conserved. Therefore, the measured resistivity $\rho$ always contains a width-independent Drude offset $\rho_0$ from the remaining bulk scattering and a width-dependent power-law component $\rho_1 w^{\beta}$. To analyze the power law component in more detail, we have subtracted $\rho_0$ from $\rho$ at all temperatures fitting the experimental data with $\rho = \rho_0 + \rho_1 w^{\beta}$. The obtained exponents $\beta$ as a function of temperature are subsequently cross-checked by a logarithmic analysis of $\rho - \rho_0 = \rho_1 w^{\beta}$ (Extended Data Fig. 8). $\log(\rho - \rho_0) = \log(\rho_1 w^{\beta}) = \log(\rho_1) + \beta \log(w)$ is a linear equation in $\log(w)$ with slope $\beta$, which can be obtained from the slope of linear fits to a log-log plot of $\rho$ vs. $w$. The determined $\beta$ are in perfect agreement with each other.

Employing the hydrodynamic model for $\beta = 2$, the viscosity can be calculated from $\rho_l = m^*/(e^2 n) \cdot 12\eta$.

**Additional magneto-hydrodynamic analysis**

Employing the hydrodynamic model for $\beta = 2$, $\rho_l$ can be expresses as $\rho_l(B) = m^*/(e^2 n) \cdot 12\eta(B) w^{-2}$, with $\eta(B) = \eta_0/(1+(2\tau_{er}\omega_c)^2)$. From this expression, we can extract $\tau_{er}$ as a function of temperature from the data in Extended Data Fig. 9 and calculate the viscosity $\eta = v_F^2 \tau_{er}/4$ as a function of temperature. As shown in Extended Data Fig. 10, the viscosities extracted from the width-dependent zero field data (Extended Data Fig. 8) and extracted from the field-dependent data in of Extended Data Fig. 9 are in excellent agreement. This agreement between the viscosities extracted from independent dependencies is an important cross-check of our interpretation and shows the consistency of our results.

With $v_F = l_{er}/\tau_{er}$ from the bulk analysis above, we can extract the momentum conserving length $l_e$ for each width independently and compare it to the momentum relaxinf mean free oath from the Hall analysis. As exemplarily shown in Extended Data Fig. 11 at 4 K, we find an excellent agreement between the mean free paths of the different widths. The obtained size-regime validates the application of the hydrodynamic model for the data where $\beta = 2$.

**Fabrication of the MEMS platforms for heat transport measurements**

The fabrication process of the MEMS devices (Extended Data Fig. 12) is similar to the process described by Karg et al.[27] A silicon wafer was coated with a 150 nm-thick layer of low-stress silicon nitride (SiMat, Germany). Gold lines with Cr adhesion layer were patterned using optical lithography, metal evaporation and a standard lift-off process. The under-etched regions were defined using optical lithography and etching of the silicon nitride. Finally, the devices were released using wet etching and critical-point drying.

The devices consist of two MEMS platforms each carrying a four-probe resistor of 430 Ω and 700 Ω, respectively, and two electrical leads to contact the sample, such that the WP$_2$ sample can be measured electrically using a four-probe geometry. Each platform is connected to the wafer via four 1.2 mm-long silicon nitride suspension legs, two of which carry three gold leads each. Two nitride bridges connecting the two platforms served to stabilize the device while placing the sample and were cut using a focussed ion beam before the measurement. The thermal conductance of each platform was $1.6 \times 10^{-5}$ W/K at room temperature and showed the expected slight increase in conductance at lower temperatures caused by the temperature-dependent thermal conductivity of silicon nitride and gold. The dimensions of the platform lead to a thermal conductance of the coupling between each heater/sensor platform to the chip carrier of $1.6 \times 10^{-5}$ W/K at room temperature. This value is 50 times larger than the sample thermal conductance of the WP$_2$ sample of $3 \times 10^{-7}$ W/K at room temperature.

**Sample Mounting on MEMS platform**

The sample was mounted using a micro-manipulator under an optical microscope. After positioning the sample, electrical contacts were deposited using electron-beam induced metal (Pt) deposition. Care was taken not to expose the WP$_2$ sample to the electron beam at any point in time except in the contact region during metal deposition. After successful mounting, the WP$_2$ sample bridges the gap between the two platforms, each equipped with a heater/sensor and two electrical leads to the WP$_2$ sample. We note that in favour of large electrical contacts and thereby good thermal coupling, the pairs of electrodes at either end of the WP$_2$ sample are not separated to allow subtracting the electrical contact resistance. However, the significant electrical resistance of the long micro-fabricated leads can be determined and subtracted from the measurement. Given the electrical conductance observed, we estimate a potential systematic error due to electrical contact resistance to be within the scatter of the experiments.

**Considerations of Thermal Contact Resistance**

To minimize the influence of thermal contact resistance, we fabricated electrical contacts to the sample with a contact area to the metal of 16 $\mu m^2$. For the phonon contribution of the thermal boundary conductance we expect values of the order of $10^{-8}$ to $10^{-7}$ $Km^2/W$. Therefore, the thermal resistance of the contacts will be on the order of $10^4 – 10^5$ K/W, which is small compared to the measured overall resistance of 1 to $4 \times 10^6$ K/W. Moreover, if we include the electron contribution to thermal conductance into this consideration, we expect the difference will be an order of magnitude further apart. Note, that although $WP_2$ is classified as a semimetal, we have an additional band at the Fermi energy contributing to the sample carrier density. The large contact size, however, increases the systematic error in the sample dimension length used in the analysis.

**Thermal Transport Measurements**

The method employed in this study is based on the method described by Li Shi et al.[32], and refined since. In short, two metal resistors are fabricated using lithography. They serve as both micro-heaters and thermometers. The resistors are read out using the four-probe technique and are fabricated on structured silicon nitride membranes to avoid thermal cross-talk via heat conduction through the substrate. The measurements are performed in vacuum to avoid thermal cross-talk through air conduction.

In detail, we proceeded as follows. First, the temperature calibration step was performed. For this, the sample was settled to set temperatures between 4 K and 300 K. The resistor on each platform was characterized measuring the voltage drop for a given current for currents between -5 and 5x10$^{-4}$ A in increments of 2 µA, in DC, from 5 to 40 mA, in AC using phase-sensitive detection modulated at 6 Hz. The electrical resistance extrapolated to the limit of zero current was used for calibration (Extended Data Fig. 13). The values measured using DC and AC operation match within the measurement error, confirming the correct choice of modulation frequency. The resulting resistance versus temperature plot shown in Extended Data Fig. 5 is used for temperature calibration of the two sensors operated as resistive thermometers.

For the subsequent transport measurements, polynomial fits for the resistance-versus-temperature for the thermometers were used; for temperatures above 50 K we use a 2$^{nd}$-order polynomial fit, for the lower temperatures we use a 5$^{th}$-order polynomial fit. At lower temperatures, the resistance versus temperature plot goes through a minimum at around 10.4 K. This is the expected behavior of gold at these temperatures [33]. As a consequence, the sensitivity is small at around this temperature and, consequently the systematic error is larger compared to high temperatures. The error bars plotted in Fig. 2 (b) of the main manuscript comprise this variation in sensitivity, as well as known systematic errors and measurement uncertainties.

After the temperature calibration, heat transport measurements were performed. At each cryostat temperature, a small sensing current of 10 nA was applied to heater/sensor 2, while the current was ramped as described above in heater/sensor 1. At all times, the current through the WP$_2$ sample was zero and the leads floating. Both sides were probed using phase sensitive detection.

For the analysis, we follow the method described by Shi et al. [32] and refined by Karg et al. [27] An example is given in Extended Data Fig. 5 for the cryostat temperature of 100 K. In summary, for each current the electrical resistance in the heater and the long leads on the suspension legs is determined. Then, using the current and the resistances, the effective Joule heating power $P_{eff}$ is calculated. From the measured resistance values $R_{Heater1}$ and $R_{Heater2}$, (Extended Data Fig. 14 (a), and (b)), the temperature rise, $\Delta T_1$ and $\Delta T_2$ of both heater/sensors is calculated (Extended Data Fig. 14 (c), and (d), respectively) using the temperature calibration (Extended Data Fig. 13). The thermal conductance of the MEMS device and the WP$_2$ sample is then extracted from the slope of plots like Extended Data Fig. 14 (c), and (d). The same results were obtained using the symmetry check at one temperature where the roles of heater and sensor side was swapped, which reconfirms the treatment of the two sensors with different resistances.

The uncertainty of the thermal conductance measurements is calculated considering known systematic errors and measurements uncertainties, the largest contributions coming from the effects of device and sample geometry, electrical resistance measurement of the sensors and the fitting procedure. The sensitivity of electrical resistance of the metal thermometers is strongly reduced at low temperatures. At low temperatures, this is the main contribution to the reported error.

**Additional key quantities**

To guide future interpretations of the data, we have calculated additional key quantities that are connected to the strength of the electron-electron interaction. In Extended Data Fig. 15 and Extended Data Fig. 16, we plot the ratio of dynamic viscosity and number density ($\eta_D$/n) in units of $\hbar$ as a function of temperature and magnetic field, respectively. $\eta_D/n$ is directly related to the momentum diffusivity.

Also, $l_{mr}k_F$ is a useful dimensionless quantity that characterizes the strength of interactions. $l_{mr}k_F > 1$ clearly supports the existence of quasiparticles in WP$_2$, in agreement with the observed Shubnikov-de Haas oscillations in the bulk samples[24].

**EXTENDED DATA**

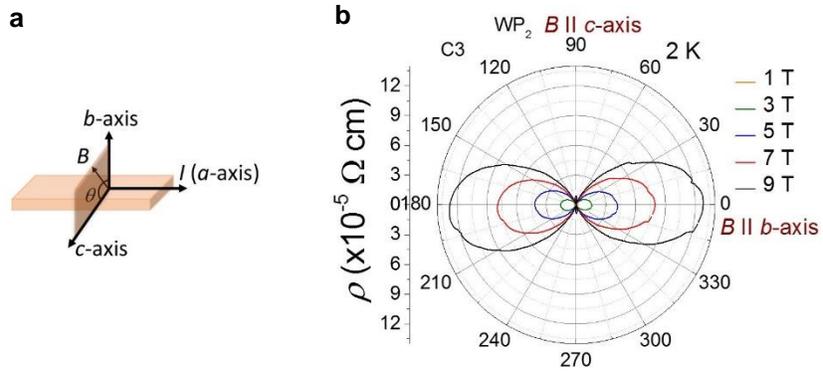

**Extended Data Figure 1 | Anisotropy of the bulk magneto-transport. a,** Sketch of the transport configuration. The measurement current $I$ is employed along the $a$-axis of the crystal. The magnetic field $B$ is applied perpendicular to the current within the $b$-$c$ plane. **b,** Anisotropy of the magnetoresistance (Ref. 24). The magnetoresistance is maximum when $B$ is parallel to $b$ and minimum when it is parallel to $c$.

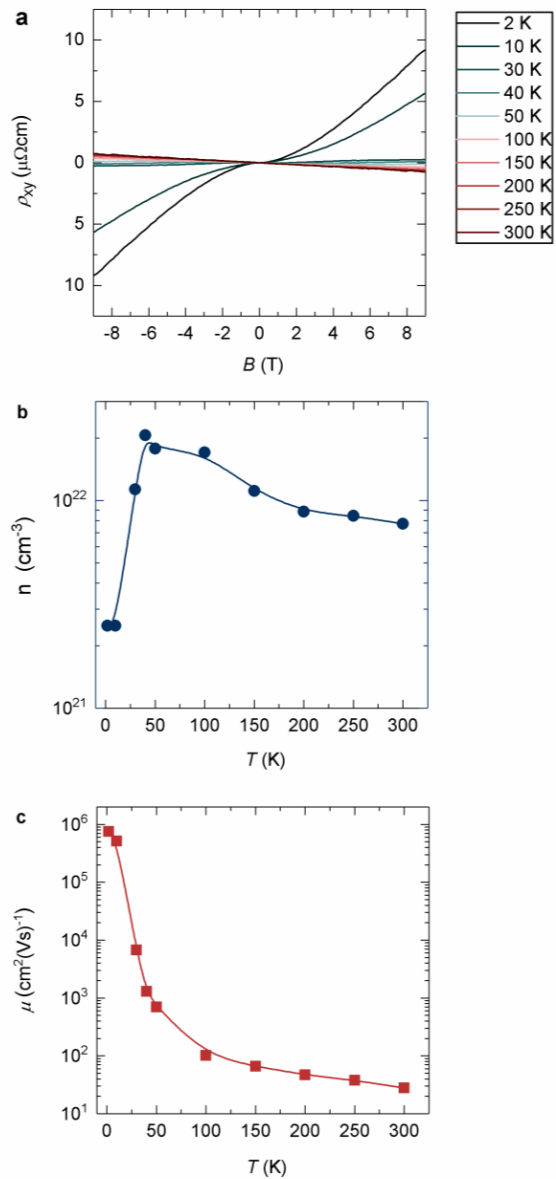

**Extended Data Figure 2 | Hall measurements on the bulk samples. a**, Hall resistivity $\rho_{xy}$ as a function of magnetic field $B$ at various temperatures. **b**, Extracted total carrier concentration $n$ and **c**, average mobility $\mu =$ as a function of temperature. $n$ is decreasing with increasing temperature due to electron-hole compensation[24].

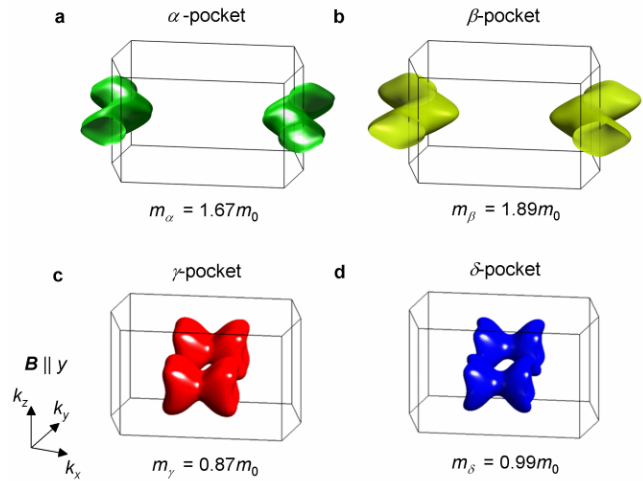

**Extended Data Figure 3 | Fermi surface and effective mass calculations** of **a**, the $\alpha$-pocket, **b**, the $\beta$-pocket, **c**, the $\gamma$-pocket and **d**, the $\delta$-pocket, for a magnetic field $B$ applied along the $b$-axis of the WP$_2$ crystal (here $y$-axis).

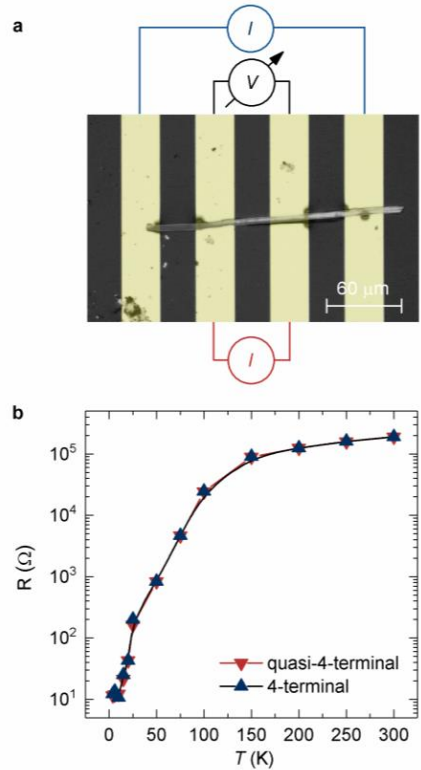

**Extended Data Figure 4 | Four-point versus quasi-four-point measurements. a,** Sketch of the transport configuration. The bias current $I$ is employed along the $a$-axis of the crystal in a quasi-4-terminal configuration at the inner contacts (red) and in an actual 4-terminal configuration (blue) at the outer contacts. The voltage response $V$ is measured at the inner contacts. **b**, Quasi-4-terminal (red) and actual 4-terminal (blue) resistivity of the 9 μm-wide WP$_2$ ribbon as a function of temperature. The deviation between the two curves is below 1 % at 300 K and rises to 10 % at 4 K. We note, however, that this enhancement at low temperature is within the measurement error, due to the low resistivity in this temperature range.

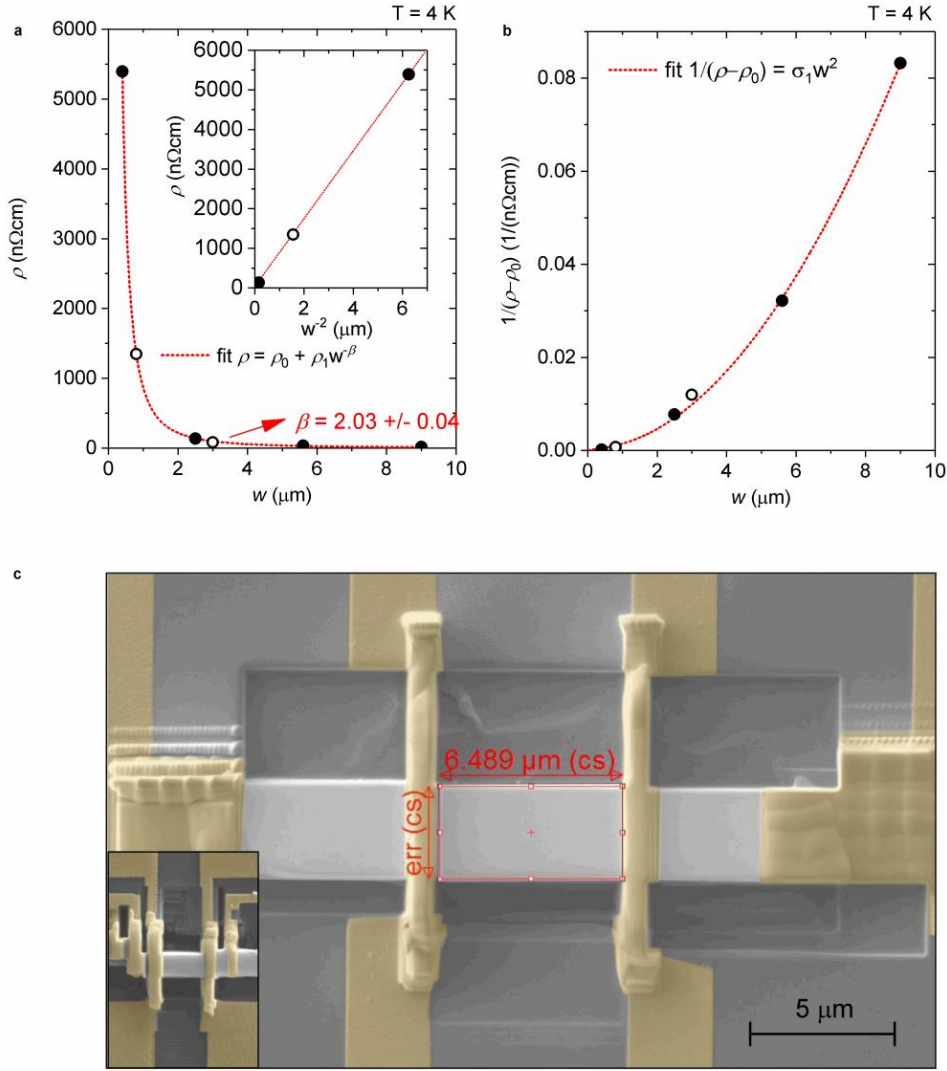

**Extended Data Figure 5 | Width-dependent electrical resistivity fits**. **a**, Measurements data (black dots) of the resistivity $\rho$ versus width $w$ at 4 K, fitted (red dashed line) by $\rho = \rho_0 + \rho_1 w^\beta$. We extract the power of $\beta = 1.96 \pm 0.03$ and a residual resistivity of $\rho_0(4\ \mathrm{K}) = 4.1 \pm 0.4$ n$\Omega$cm. The inset shows the measurement data and fit of $\rho$ versus $w^{-2}$. **b**, $1/(\rho-\rho_0)$ versus $w$, showing the $w^2$-dpendence of the viscous conductance at 4 K. **c**, Scanning electron microscope image of the four-terminal devices, cut by a Focused Ion Beam. The scale bar counts for both images.

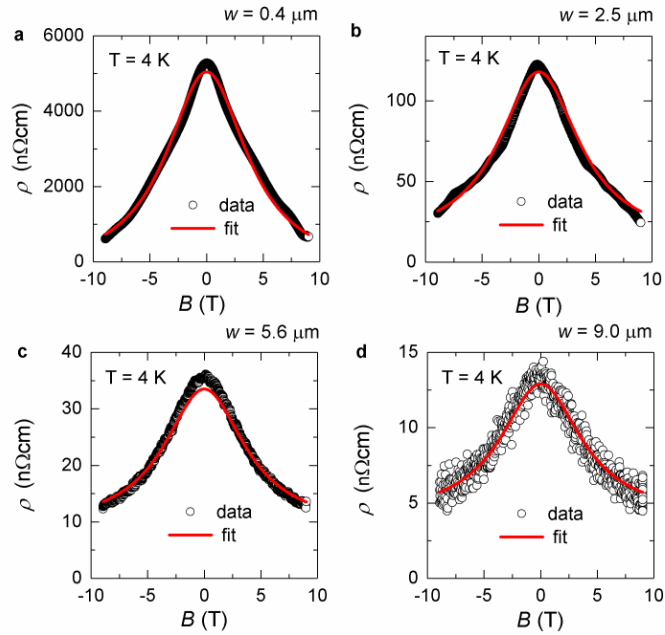

**Extended Data Figure 6 | Magneto-resistivity at 4 K** of **a**, the 0.4 μm-wide sample, **b**, the 2.5 μm-wide sample, **c**, the 5.6 μm-wide sample and the **d**, the 9.0 μm-wide sample. The black circles denote the measurements data and the red lines are best fits of $\rho = \rho_0 + \rho_{1,a} w^{\beta}/(1+(\rho_{1,b}B)^2)$, from which the residual resistivity $\rho_0$ can be extracted for each width individually.

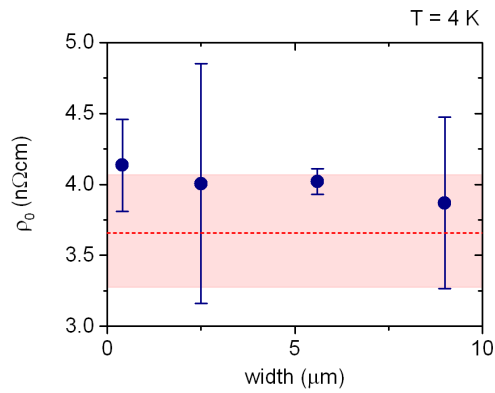

**Extended Data Figure 7 | Comparison of the residual resistivity $\rho_0$ at 4 K**, extracted from the zero-field width-dependent electrical resistivity (red dashed line, the light red area denotes the error from fit) shown in Extended Data Fig. 5 and from the magneto-resistivity (blue dots, error bars denote the errors from the fits) shown in Extended Data Fig. 6. The differently extracted $\rho_0$ are similar, demonstrating the consistency of our results and analysis.

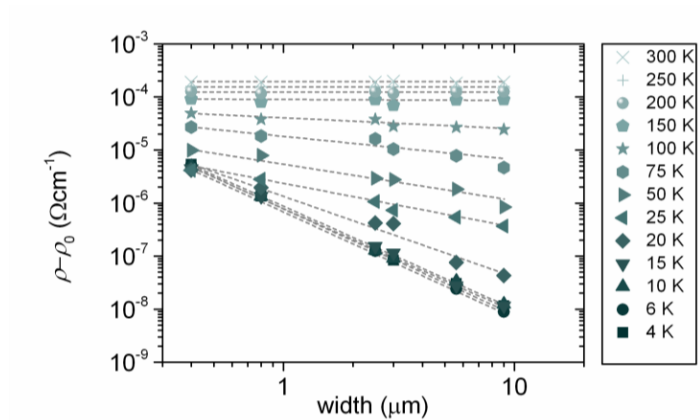

**Extended Data Figure 8 | Extraction of the exponent of the functional dependence of $\rho-\rho_0$ on *w* at various temperatures.** The slope of the linear fits (dashed lines) from the experimental log-log data (symbols) is plotted in Fig. 1 (d).

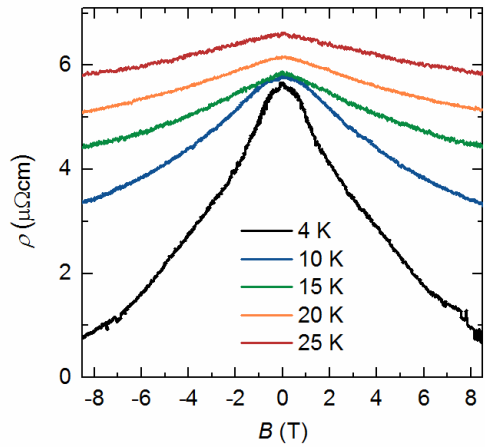

**Extended Data Figure 9 | Electrical resistivity $\rho$ of the 2.5 μm-wide beam as a function of magnetic field $B$ at various temperatures.**

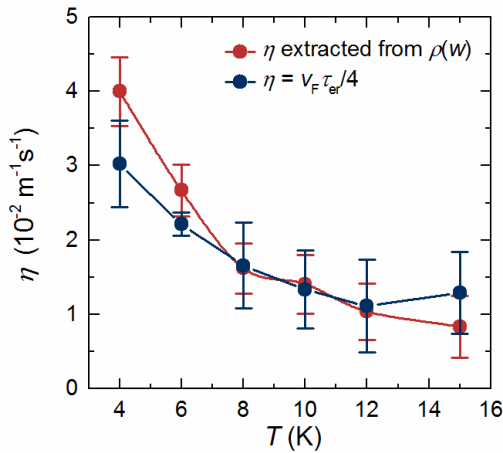

**Extended Data Figure 10 | Kinematic shear viscosity $\eta$ versus temperature.** The data is extracted from the functional dependence $\rho \sim w^{-2}$ (Extended Data Fig. 8) (red dots) and from the relaxation times $\tau_{er}$ as $\eta = v_F^2 \tau_{er}/4$.

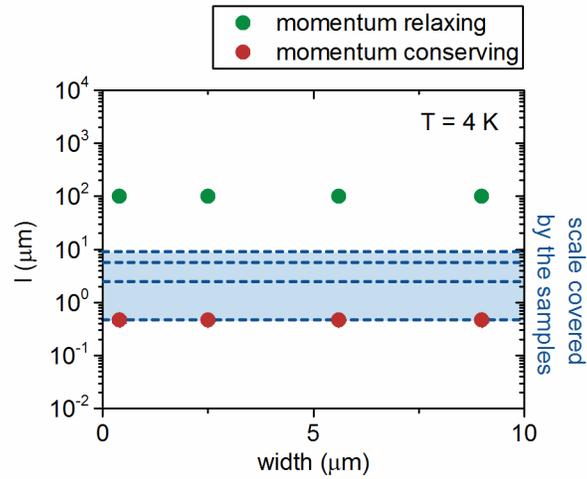

**Extended Data Figure 11 | Momentum-relaxing (green dots) and momentum-conserving (red dots) scattering length at 4 K**, obtained from individual fits of the data in Extended Data Fig. 7 by the Navier-Stokes flow model. The blue dashed lines denote the width of the individual measured samples. The light blue area shows the investigated size-range that lies well within the hydrodynamic regime at the boundary to ballistic crossover. However, we do not observe signatures for ballistic transport in our experiments.

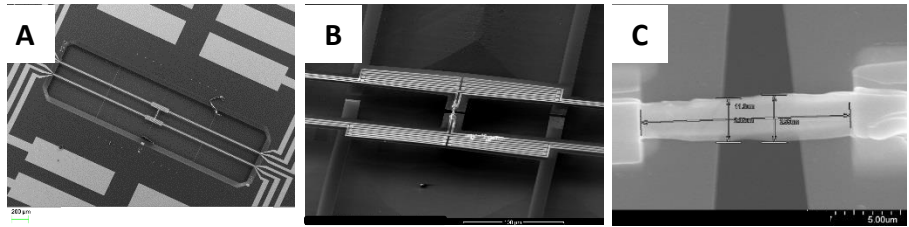

**Extended Data Figure 12 | The MEMS platform for thermal measurements. a**, Scanning electron micrograph of measurement device similar to the one used in this study prior to mounting the sample. The 1.2 mm-long MEMS beams are etched out of silicon nitride and carry three gold lines each. **b**, Enlarged area of the measurement device with mounted $WP_2$ sample. The stripes for stabilization the structure during fabrication (connecting the two heater areas) were cut using a focused ion beam. **c,** Enlarged image of sample piece. Note that the $WP_2$ sample is in mechanical contact only with the gold and the deposited Pt lines, but not with the nitride membrane that forms a V-shaped gap.

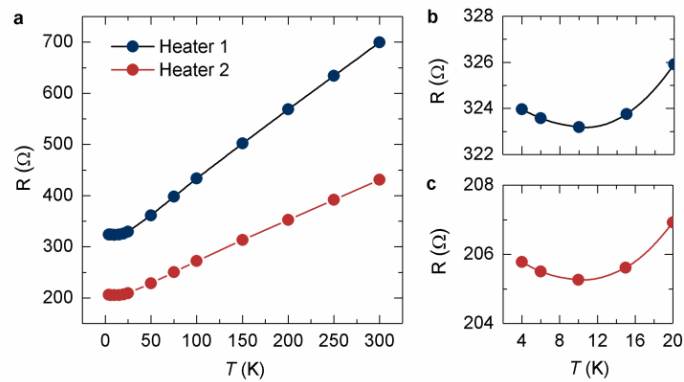

**Extended Data Figure 13 | Temperature calibration plot of the two heater/sensors of the measurement device. a,** The plot shows the data of heater 1 and heater 2 (extrapolated resistance to zero drive current) as dots and the calibration fits as lines versus cryostat temperature. **b**, and **c**, are zoom-ins at low temperatures for heater 1 and 2, respectively.

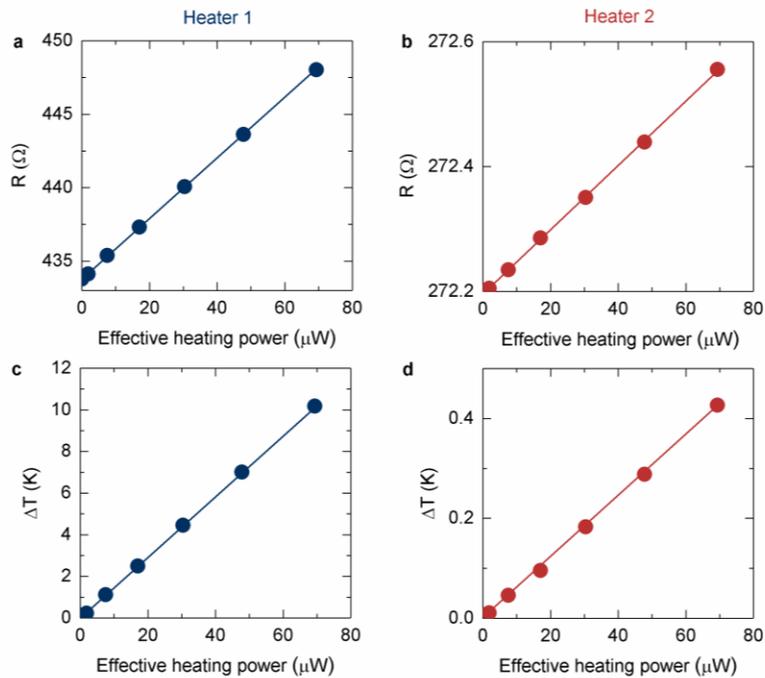

**Extended Data Figure 14 | Example of the analysis of the data used to calculate the thermal conductance of the WP$_2$ sample.** Here data for a cryostat temperature of 100 K is shown. Resistance of heater **a**, and sensor **b**, for the given heating and sensing currents. Temperature rise above 100 K for heater **c**, and sensor **d**.

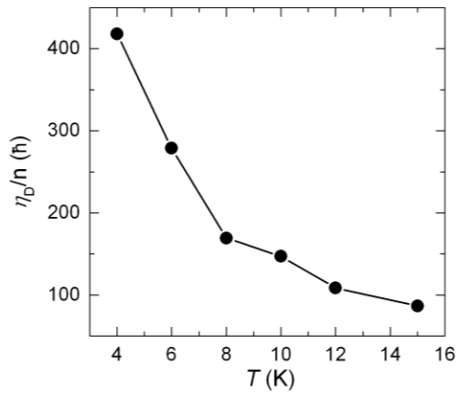

**Extended Data Figure 15 | Ratio of dynamic viscosity and number density ($\eta_D/n$) in units of $\hbar$ as a function of temperature.**

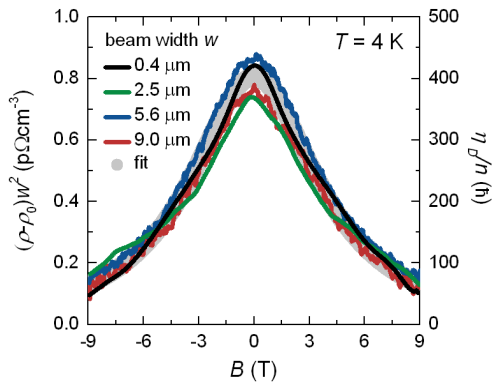

**Extended Data Figure 16 | Ratio of dynamic viscosity and number density ($\eta_D/n$) in units of $\hbar$ as a function of magnetic field at 4 K (right axis). Left axis: $(\rho-\rho_0)w^2$.**

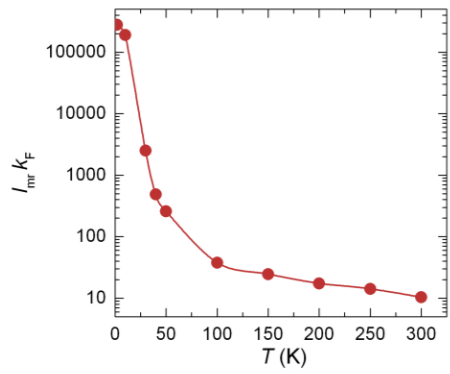

**Extended Data Figure 17 | $l_{mr}k_F$ as a function of temperature.** $l_{mr}k_F > 1$ clearly supports the existence of quasiparticles in WP$_2$.

**ACKNOWLEDGEMENTS**

The authors thank Steffen Reith and Valentina Troncale for technical support, as well as Stuart Parkin, Philip Moll, Subir Sachdev, Sean Hartnoll, Jan Zaanen and Andrew Mackenzie for fruitful discussions. We also acknowledge support by Walter Riess and Kirsten Moselund, and thank Charlotte Bolliger for copy-editing. Fabian Menges gratefully acknowledges the support from the Society in Science – The Branco Weiss Fellowship and a Swiss National Science Foundation postdoctoral fellowship

## AUTHOR INFORMATION

### Contributions

J. G., F. M., C. F. and B. G. conceived the experiment. C. S., N. K. and V. S. synthesized the single-crystal bulk samples. U. D. fabricated the suspended platforms. J. G., F. M. and B.G. produced the micro-ribbons and processed the devices. J. G. carried out the transport measurements on the micro-ribbons with the help of R. Z. J. G. and B. G. analyzed the data. C. S. and N. K. performed the Hall and resistivity measurements on the bulk samples. Y. S. calculated the band structure. B. G. supervised the project. All authors contributed to the interpretation of the data and to the writing of the manuscript.

### Data availability statement

All data generated or analysed during this study are included in this published article (and its supplementary information). The datasets generated during and/or analysed during the current study are available from the corresponding author on reasonable.

### Competing financial interest

The authors declare no competing financial interests.

### Corresponding author

* johannes.gooth@cpfs.mpg.de, bgo@zurich.ibm.com